\documentclass[conference,letterpaper]{IEEEtran}
\usepackage{cite}
\usepackage{amsmath,amssymb,amsfonts}
\usepackage{textcomp}

\usepackage{cmap}
\usepackage[T1]{fontenc}

\usepackage{graphicx}

\usepackage{array}

\usepackage{multirow}

\usepackage[misc]{ifsym}

\usepackage{bbding}

\usepackage{csquotes}
\usepackage{url}
\usepackage{paralist}
\usepackage[draft]{hyperref}

\makeatletter
\newcommand{\rmnum}[1]{\romannumeral #1}
\newcommand{\Rmnum}[1]{\expandafter\@slowromancap\romannumeral #1@}
\makeatother

\begin{document}
	
	\title{CEFIoT: A Fault-Tolerant IoT Architecture for Edge and Cloud}
	
	\author{\IEEEauthorblockN{Asad Javed\IEEEauthorrefmark{1}, Keijo Heljanko\IEEEauthorrefmark{1}\IEEEauthorrefmark{2}, Andrea Buda\IEEEauthorrefmark{1} and Kary Fr{\"a}mling\IEEEauthorrefmark{1}}
		\IEEEauthorblockA{\IEEEauthorrefmark{1}Department of Computer Science, Aalto University, Espoo, Finland \\
			\IEEEauthorrefmark{2}{Helsinki Institute for Information Technology HIIT} \\
			\{asad.javed, keijo.heljanko, andrea.buda, kary.framling\}@aalto.fi}
	}

	\IEEEoverridecommandlockouts
	\IEEEpubid{\makebox[\columnwidth]{978-1-4673-9944-9/18/\$31.00 \copyright~2018~IEEE}
	\hspace{\columnsep}\makebox[\columnwidth]{ }}
	
	\maketitle
	
	\begin{abstract}
	Internet of Things (IoT), the emerging computing infrastructure that refers to the networked interconnection of physical objects, incorporates a plethora of digital systems that are being developed by means of a large number of applications. Many of these applications administer data collection on the edge and offer data storage and analytics capabilities in the cloud. This raises the following problems: (\rmnum{1})~the processing stages in IoT applications need to have separate implementations for both the edge and the cloud, (\rmnum{2}) the placement of computation is inflexible with separate software stacks, as the optimal deployment decisions need to be made at runtime, and (\rmnum{3})~unified fault tolerance is essential in case of intermittent long-distance network connectivity problems, malicious harming of edge devices, or harsh environments. This paper proposes a novel fault-tolerant architecture~CEFIoT for IoT applications by adopting state-of-the-art cloud technologies and deploying them also for edge computing. We solve the data fault tolerance issue by exploiting the Apache Kafka publish/subscribe platform as the unified high-performance data replication solution offering a common software stack for both the edge and the cloud. We also deploy Kubernetes for fault-tolerant management and the advanced functionality allowing on-the-fly automatic reconfiguration of the processing pipeline to handle both hardware and network connectivity based failures. 
	\end{abstract}
	
	
	\section{Introduction}	
	Internet of Things (IoT) is becoming a promising paradigm for the future Internet \cite{DBLP:journals/cn/AtzoriIM10}. It provides a network where information flows could easily be set up between any kind of products, devices, users, and information systems in general. The key goal of IoT is to link real-world objects with the virtual world, thus enabling connectivity anywhere, anytime, and with anything \cite{DBLP:journals/percom/KublerFB15}. This vision is becoming real due to the continuous development of new information system concepts and technologies which incorporate a large number of similar and heterogeneous applications. 
	
	Many of these applications administer data collection on the edge and offer data storage and analytics capabilities in the cloud. It is often necessary to move computation such as alarm generation from raw data streams to the edge, as it saves a large amount of network bandwidth that might be unavailable for all the sensors. In some other cases, ample network bandwidth can be available for nodes with better connectivity but the computing capabilities at edge might be insufficient. In such cases, some processing steps might need to be transferred to the cloud from the edge. Such compute placement decisions are dependent on many issues and the optimal deployment decisions need to be made at runtime. However, this raises a problem where the processing stages in IoT applications need to have two implementations: one for the edge and one for the cloud. 
	
	This brings us to the other known challenges in the existing IoT systems. (\rmnum{1})~A common software stack is necessary for computing, portability, and management ease that allows data processing to be moved between edge and cloud. (\rmnum{2})~In the scenario where edge nodes interact with a specific cloud back-end, the available network bandwidth between the edge and the cloud becomes a bottleneck for large data transportation. This requires computing to be performed at the edge nodes in order to minimize physical distance delay and consume a smaller amount of bandwidth~\cite{GS:rabinovich2004computing}. (\rmnum{3})~In a clustered system of many nodes where data are transported between edge and cloud, it might be possible to lose the data forever due to malfunction of a single edge node. Local fault tolerance needs to be implemented to preserve the system state locally at the edge especially in the case of a node failure or intermittent long-distance network connectivity problems. 
	
	To address these challenges, this paper proposes CEFIoT, a new fault-tolerant architecture for IoT applications by adopting state-of-the-art cloud technologies and deploying them also for edge computing. CEFIoT architecture is composed of three layers: (\rmnum{1})~Application Isolation, (\rmnum{2})~Data Transport, and (\rmnum{3})~Multi-cluster Management layer. Based on this layered design, the architecture allows compute placement on either the edge or the cloud without source code modifications. This is enabled by using the same lightweight Docker container-based software stack on the cloud and on the edge, and by deploying embedded Linux devices at the edge. We consider a class of applications, such as the surveillance camera demonstrator system we have built, where the edge also has capabilities to operate under hardware faults (such as malicious harming of edge devices or harsh environments). This requires fault tolerance using replication of data to be processed on several edge devices, reconfiguring the data processing pipeline when hardware or network failures occur, and capabilities to operate independently on the edge in a degraded fashion when disconnected from the cloud back-end. In this way, the same fault tolerance capabilities of CEFIoT can be applied to other hostile environments such as vehicle control system and ships that may have only intermittent network connectivity.
	
	All these issues of fault tolerance are addressed in CEFIoT by employing a small cluster of embedded Linux devices (e.g. Raspberry Pi (RPi) nodes in our demonstrator) together with Kubernetes and Apache Kafka to implement the fault tolerance capabilities. We solve the data fault tolerance issue by adopting the Kafka publish/subscribe (pub/sub) platform as the unified high-performance data replication solution for both edge and cloud. We also deploy Kubernetes for management and the advanced functionality of allowing on-the-fly automatic reconfiguration of the processing pipeline to handle both hardware and network connectivity based failures. Our pipeline is fully fault-tolerant in the sense that the failure of any single computing node on either the edge or the cloud will not disrupt the data processing pipeline, but the architecture will self-adapt and reconfigure around any single node failure. This is the case also for the management nodes, which are replicated using Kubernetes fault tolerance functionality.
	
	The rest of the article is organized as follows. Section \ref{sec:related} presents the related work in IoT architectures. In Section \ref{sec:design}, the layered fault-tolerant architecture CEFIoT is proposed and elaborated. Section \ref{sec:casestudy} describes a surveillance camera case study in which the fault tolerance capability of CEFIoT is demonstrated. Finally, Section \ref{sec:conclusion} concludes this paper. 
	
	\section{Related Work}\label{sec:related}
	Due to the rapid expansion of IoT, demand for computing and communication devices in network-based infrastructure has also increased. However, one fundamental principle emerges: the greater the benefits these devices provide to the well-being, the higher the potential for harm when they are unable to function correctly. In that case, fault tolerance is the best guarantee as it is capable of overcoming the physical, design, or human-machine interaction faults and to preserve the correct execution of the tasks \cite{DBLP:journals/computer/Avizienis97}\cite{DBLP:journals/tc/Avizinis76}. 
	
	Typically, in the real world, building a fault-tolerant system for IoT is a complex task, mainly because of the extremely large variety of edge devices, data computing technologies, networks, and other resources that may be involved in the development process \cite{DBLP:journals/iotj/ZanellaBCVZ14}. To accomplish such a system, virtualization technologies have been designed. Container-based virtualization \cite{DBLP:conf/eurosys/SolteszPFBP07}\cite{DBLP:conf/ic2e/MorabitoKK15} has become the apparent choice in terms of performance as compared to the traditional hypervisor-based virtualization \cite{DBLP:conf/eurosys/SolteszPFBP07}\cite{DBLP:conf/icppw/HuaiLH07}. Through containers and the use of high-level languages and a common software stack, the same analytics programs can be executed without source code modifications at the edge. 	 
	
	Academic and industrial research abounds in literature exploring containers and fault-tolerant IoT architectures. However, to the best of our knowledge, this work is the first to develop an architecture which provides both node and network fault tolerance for the edge and the cloud using the same software framework for both. The design also takes into consideration the limited resources available at the edge by employing lightweight containers instead of using traditional virtual machines. We discuss some of the recent work on designing an IoT applications architecture.
	
	A platform known as resin.io\footnote{\url{https://resin.io/how-it-works/}} has been developed which uses Docker-based management, instead of Kubernetes cluster orchestration, for IoT devices. The purpose of this platform is to introduce the benefits of Linux containers to IoT for deploying and managing isolated applications. Although the platform is robust to sudden power failures and disk corruption, it has no functionality for managing several embedded devices as a single fault-tolerant cluster. There are other well-known platforms such as Azure IoT Suite, Google Cloud IoT, and Amazon AWS, which deliver fully integrated cloud services and allow many systems to easily connect, manage, and ingest IoT data on a large scale. However, these platforms provide no edge fault tolerance.  
	
	Netto et al. \cite{DBLP:journals/jsa/NettoLCLS17} propose the integration of coordination services by developing a state machine replication system based on Docker containers that is managed by Kubernetes. Krco~et~al. \cite{DBLP:conf/wf-iot/KrcoPC14} provide an overview of designing IoT architecture in ETSI M2M, FI-WARE, IoT6, and IoT-A projects combined with cloud computing capabilities.
	
	Gia et al. \cite{GS:gia2015fault} propose a fault-tolerant and scalable IoT architecture for health care which covers many fault situations including malfunction of sink node hardware and traffic bottlenecks. Similarly, another fault-tolerant mechanism for intelligent IoT is implemented by Su et al. \cite{DBLP:conf/wf-iot/SuSHLW14} that presents the design of a fail recovery mechanism in WuKong middleware. These approaches provide a low-cost distributed solution and are capable of fail-over in a small network.   
	
	Belli et al. \cite{DBLP:conf/esocc/BelliCFMP14}\cite{DBLP:journals/ijssoe/BelliCDFMMP15} present a cloud architecture for the management of large stream applications that can efficiently handle real-time data through a processing platform. This architecture delivers processed data to consumer applications with low latency and no fault tolerance mechanism.
	
	\section{CEFIoT: Proposed IoT Architecture}\label{sec:design}	
\begin{figure}[t]
		\centering \includegraphics[scale=0.45]{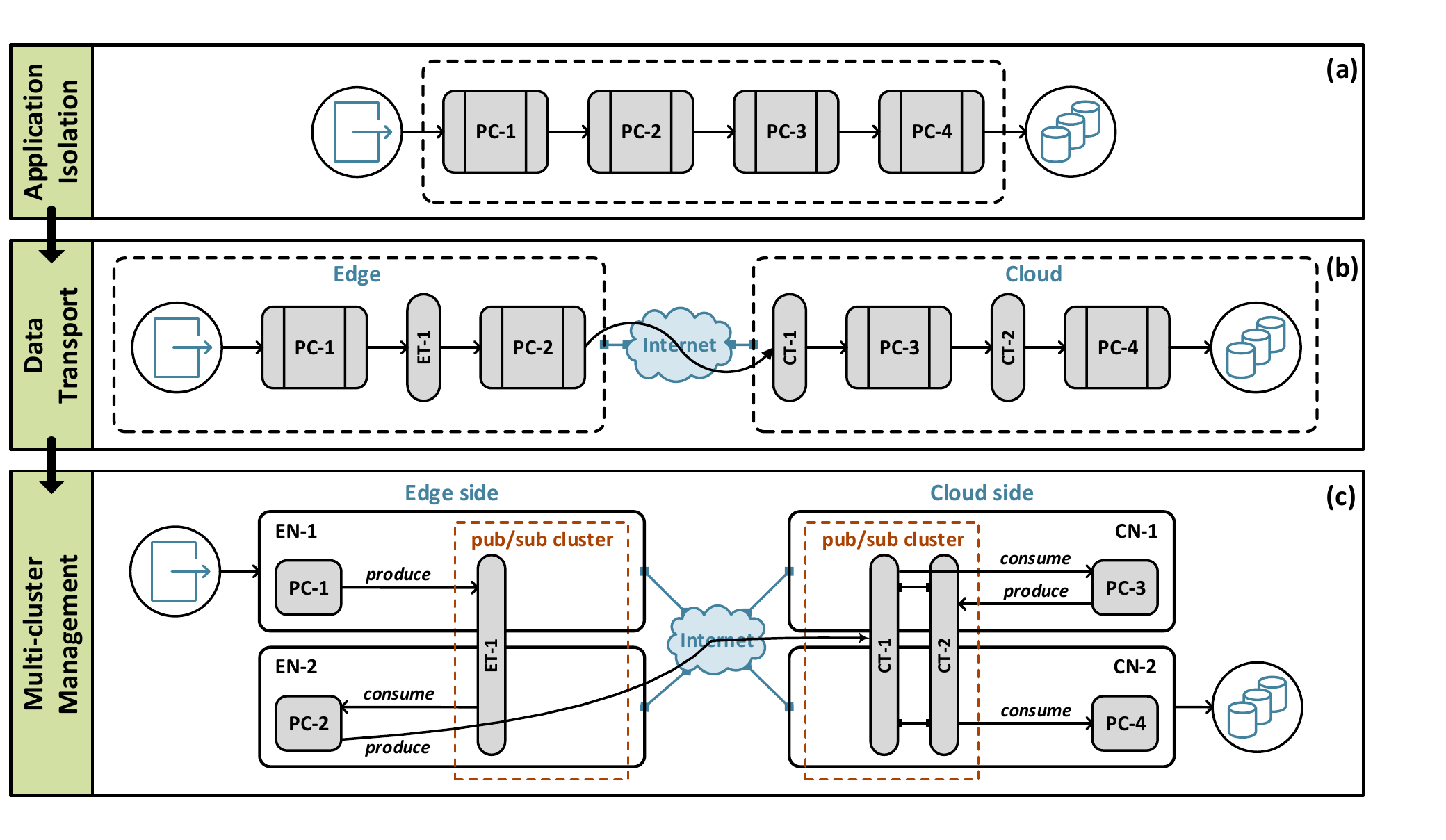}
		\caption{CEFIoT Layered architecture; where $PC$ Processing Container, $ET$ Edge Topic, $CT$ Cloud Topic, $EN$ Edge Node, $CN$ Cloud Node}
		\label{fig:larch}
	\end{figure}

	The CEFIoT architecture is designed to offer unified multi-cluster management which provides fault tolerance for both the edge- and the cloud-side clusters and is capable of fail-over in a large interconnected network. In addition, separate clusters for edge and cloud in CEFIoT allow edge devices to operate independently also when disconnected from the cloud back-end. Thus, the architecture operates in a degraded mode even when cloud connectivity is lost. This architecture is logically constructed from the underlying technologies with three layers of abstraction called 
	\begin{inparaenum}[1)]
		\item Application Isolation,
		\item Data Transport, and
		\item Multi-cluster Management layer,
	\end{inparaenum}
	as shown in Fig.~\ref{fig:larch}. In the running example depicted in Fig.~\ref{fig:larch}, we have an application with four processing stages: \textit{PC-1, PC-2, PC-3,} and \textit{PC-4}.
	
	\begin{enumerate}
		\item The \textbf{Application Isolation} layer of CEFIoT wraps individual processes into separate independent blocks and configures them to operate as a single isolated application for transporting data from source to destination (see Fig.~\ref{fig:larch}(a)). This wrapping is achieved by adopting Linux containers that provide application isolation for convenient software deployment and also allow the processing of data streams irrespective of the physical hardware. In the case of machines with different architecture types, containers execute the same application-level code (such as Java, Python, or Spark) without source code modifications to ensure software portability. Docker is chosen for that purpose as it is by far the most mature and adopted development tool creating distributable applications and configuring them to interact with the outside world. It also isolates several IoT applications running on the same edge and cloud nodes from each other.    
		
		\item The \textbf{Data Transport} layer of CEFIoT provides a pub/sub messaging framework in which streams of data are buffered and replicated across a cluster. This allows the architecture to have logical data flow in the form of containerized processes using pub/sub topics as the transport medium. In this way, processing can be distributed easily, which gives us location flexibility for computation to be placed on either the edge or in the cloud. This organization of data pipeline maintains network fault tolerance allowing data buffering locally at the edge, while Internet connectivity is being reconfigured. Several dedicated pub/sub topics are used to buffer data streams as depicted in Fig.~\ref{fig:larch}(b), which allows the data to be available all the time in the cluster even while some processing stages are being reconfigured.
		
		This layer functionality is achieved by adopting the Apache Kafka framework as it solves the problem of data fault tolerance by providing both on-line and off-line messages consumption \cite{GS:kreps2011kafka}. Unlike the other systems used by CEFIoT to store configuration data, Kafka is tailored for very high data rates. As an example, it is used by Netflix to collect their log data streams\footnote{\url{https://www.slideshare.net/wangxia5/netflix-kafka}}. Other queuing systems such as RabbitMQ could be considered alternatives to Kafka for providing the needed functionality. 
		
		\item The \textbf{Multi-cluster Management} layer of CEFIoT is responsible for managing and monitoring the operations of upper two layers by providing a unified management system. In this layer, the data processing is assigned to the physical machines/nodes based on the load balancing and fault tolerance requirements of processing stages. Fig.~\ref{fig:larch}(c) illustrates this placement in the form of Edge Node (EN) and Cloud Node (CN). Since the data are buffered in topics, they are also replicated to other nodes for scenarios where some system node becomes unresponsive and the data can be consumed from any other available node. In such cases, this layer overcomes the problem of node failures by re-scheduling failed processing stages on other available physical nodes.
		
		This management system is achieved through the Kubernetes framework in which numerous containers can be executed as a cluster of encapsulated applications, isolated from each other. With Kubernetes, it is possible to deploy high-availability applications, scale and manage them during runtime, and use application-specific resources during execution \cite{DBLP:journals/cloudcomp/Bernstein14b}. One of the key design principles of Kubernetes is to have configuration data replication, allowing Kubernetes to survive any single node failure. 
	\end{enumerate}
	
	The layered architecture of CEFIoT starts with the application isolation layer in which we have (in our running example) four Processing Containers (PCs) that behave as an isolated application as depicted in Fig. \ref{fig:larch}(a). These processes are connected in a sequential order, and together they become a single application that runs on a single machine and transports data from the source to the destination. 
	
\begin{figure}[t]
		\centering \includegraphics[scale=0.49]{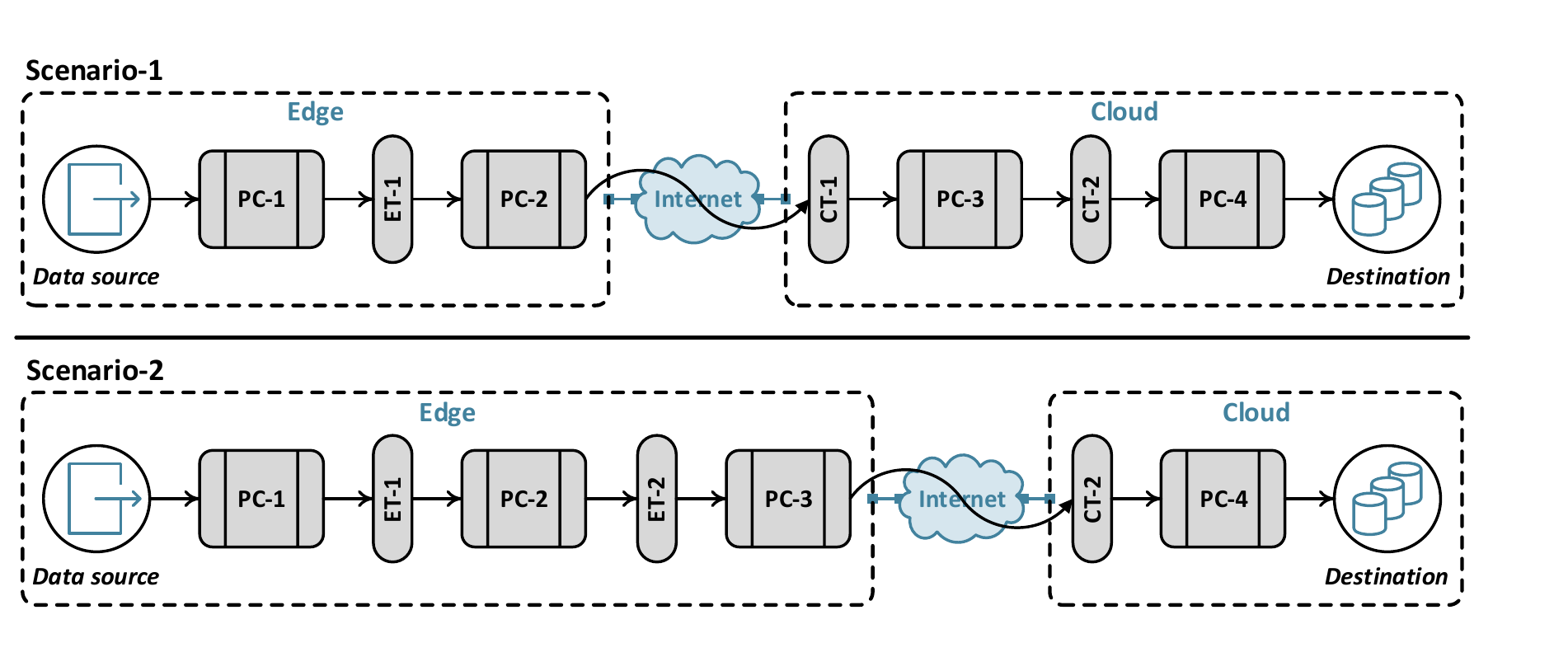}
		\caption{Logical data pipeline of CEFIoT; where $PC$ Processing Container, $ET$ Edge Topic, $CT$ Cloud Topic}
		\label{fig:garch}
	\end{figure}

	The data transport layer of CEFIoT is further extended to two scenarios (see Fig. \ref{fig:garch}), where Kafka is deployed as the pub/sub cluster to provide data stream transportation and allow containers to move between edge and cloud. As can be seen, both edge and cloud sides consist of three Kafka topics along with four PCs. This, in turn, represents a logical data flow of processing stages. In Scenario-1 (see Fig.~\ref{fig:garch}), the edge side has two separate containers, where PC-1 collects data from the data source, performs pre-processing (for instance, filtering or compression), and sends them to the local Edge~Topic~1~(ET-1). PC-2 then consumes the data from ET-1, performs additional processing, and sends them to the cloud side on Cloud~Topic~1~(CT-1). Similarly, the cloud side also contains two separate containers PC-3 and PC-4 which consume data from topics CT-1 and CT-2 respectively, process them further, and deliver the processed data to the destination. The overall data stream is deployed through separate containers along with the pub/sub topics. In this way, it is easier to place data computation either on the edge or in the cloud, as the two sides have a common software stack.
	
	Scenario-2 depicts the behavior in which PC-3 is moved to the edge side. This shows the location flexibility for data processing, since PC-3 is a separate container and both sides have similar Kafka cluster. This configuration gives us benefits in terms of minimum bandwidth consumption and less latency. Additionally, if there is a network connectivity problem, the data will always be available on the edge side. Once the Internet outage has been resolved, for example, by using a secondary network connection, the architecture resumes processing on both edge and cloud sides. In both scenarios, the logical data flow is actively monitored through Kafka which keeps the pipeline active at all times.
	
\begin{figure}[!htb]
		\centering \includegraphics[scale=0.51]{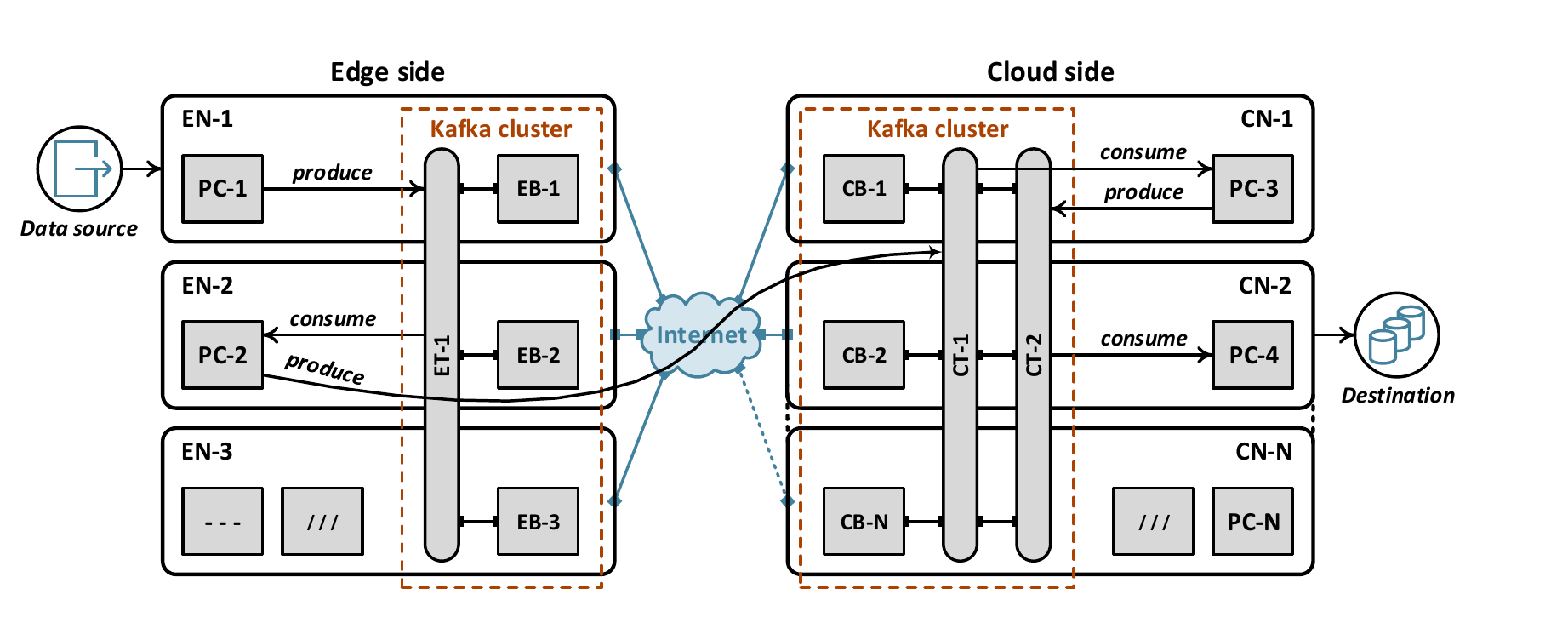}
		\caption{Mapping of Scenario-1 to physical machines; where $EB$ Edge Broker, $CB$ Cloud Broker, $EN$ Edge Node, $CN$ Cloud Node}
		\label{fig:garchkafka}
	\end{figure}

\begin{figure}[!htb]
		\centering \includegraphics[scale=0.51]{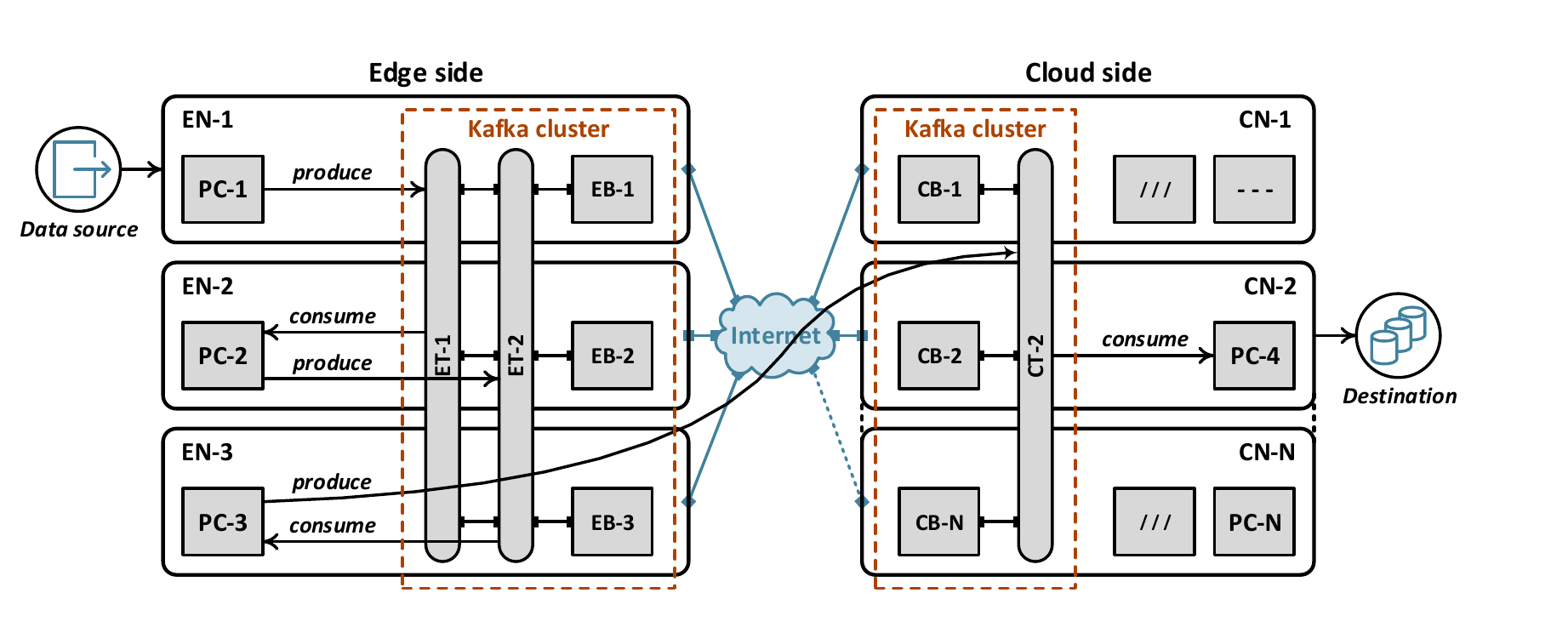}
		\caption{Mapping of Scenario-2 to physical machines}
		\label{fig:garchkafka2}
	\end{figure}

	Both scenarios of Fig. \ref{fig:garch} are then mapped to the physical machines. Fig. \ref{fig:garchkafka} and Fig. \ref{fig:garchkafka2} show the extension of CEFIoT to the multi-cluster management layer. As can be seen, the edge side contains three nodes along with the local Kafka cluster which includes multiple topics and Kafka brokers. Similarly, the cloud side has an \textit{N}~number of machines along with the remote Kafka cluster. The mapping provides a more detailed overview in which there are three brokers on the edge side, thus providing 3-way replication of data. Both sides have Kafka topics that are accessible in each physical node. In this way, if some node failures happen, the data can be consumed from any other available node. Fig.~\ref{fig:garchkafka} and Fig.~\ref{fig:garchkafka2} are basically drawn to demonstrate that (\rmnum{1}) sensors and processing can be on different edge devices, as long as they have access to the same Kafka cluster, thus providing computing location independence, (\rmnum{2}) there is a flexibility to move processing stages between edge and cloud, and (\rmnum{3}) in the case of an Internet outage, the data are always available and buffered locally. Once the network connectivity has been resolved, the data are sent from local edge Kafka to the cloud Kafka. 
	
\begin{figure}[t]
		\centering \includegraphics[scale=0.51]{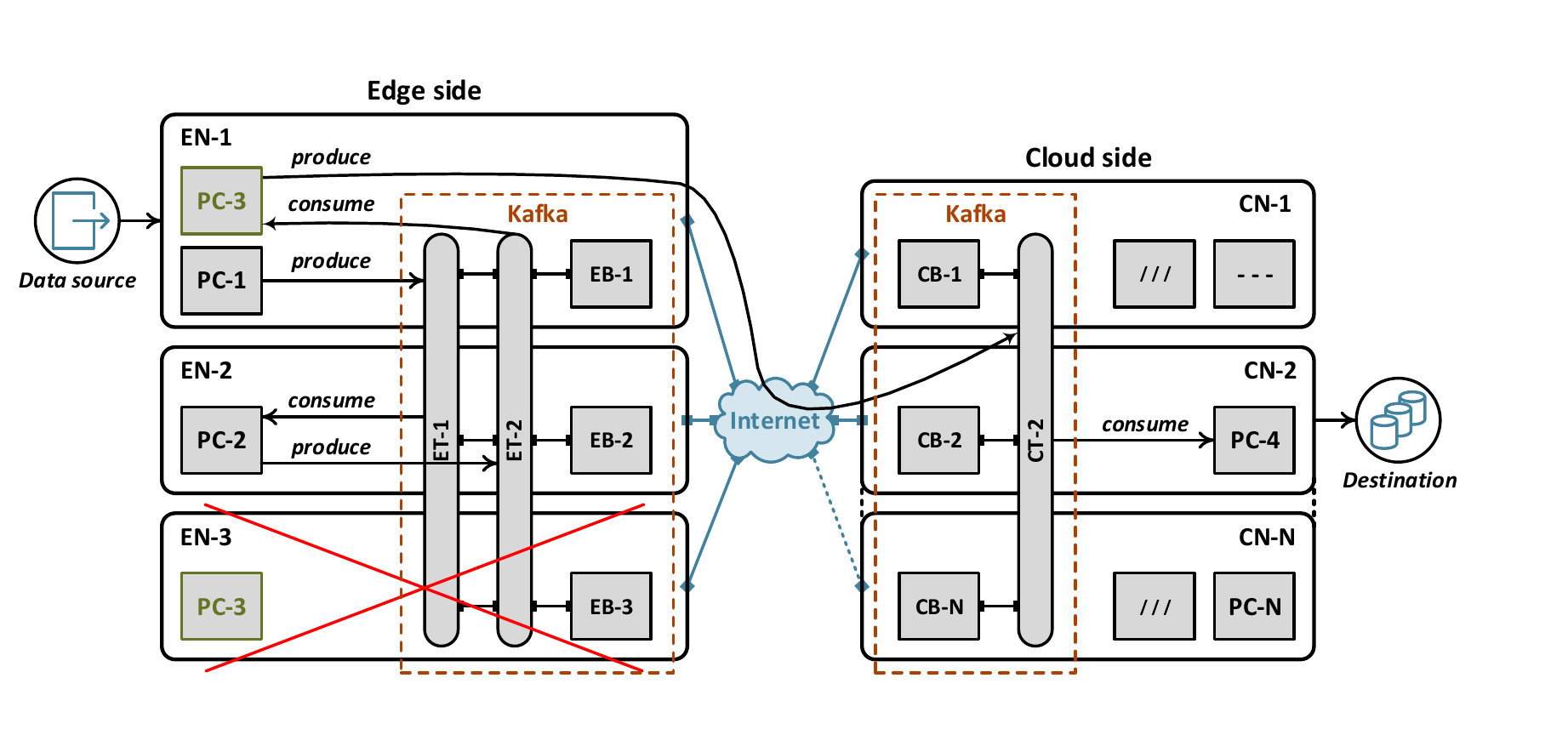}
		\caption{Scenario-3 in which edge node EN-3 fails, PC-3 is resumed on EN-1}
		\label{fig:garchkafka3}
	\end{figure}

	Fig. \ref{fig:garchkafka2} is further extended to Scenario-3 in which Kubernetes is deployed as a cluster orchestration system (see Fig.~\ref{fig:garchkafka3}). As can be seen, Edge Node 3 (EN-3) becomes unresponsive at the edge side. Kubernetes reschedules PC-3 on EN-1. Since the data reside on the topic which is replicated between all nodes, PC-3 consumes the data from ET-2 and sends to the topic CT-2 on the cloud side from node EN-1. In this way, the architecture handles node failure by maintaining the correct system state in which the physical data flow is changed; however, the logical data flow remains the same. This offers a fault tolerance functionality that is fully transparent to the application programmer and also allows for dynamic load balancing.
		
	\section{Case Study: Surveillance Camera System}\label{sec:casestudy}	 
	The layered design of CEFIoT is applied on a surveillance camera use case as an example of a generic fault-tolerant IoT application. In this case study, a surveillance camera system equipped with motion detection is demonstrated in a potentially hostile environment, where images with movements in the live video feed are captured on the edge side. This enables saving network bandwidth, processing the images by adding useful information (such as camera id, date, and time-stamp), and sending to the cloud for permanent storage. In case of a criminal activity, there is a possibility that some system nodes may be physically damaged or the network connectivity may be temporarily cut off. The system should have a data processing capability to preserve and buffer data streams locally at the edge especially in the scenario of intermittent long-distance network connectivity outage or malicious harming of devices. This requires fault tolerance to be implemented for the overall system including both the edge- and the cloud-side clusters. Through this way, the system operates in a degraded mode storing data locally at the edge even during the time cloud connectivity is lost.
	
	All these problems are addressed by constructing a model use case as a clustered system of five RPi nodes, each with a camera attached to them, together acting as a surveillance camera system. RPi is selected merely for the testing purpose as it is a fully functional ARM-based Linux edge device which allows the same Linux-based application software to run both on the edge and in the cloud. The case study implementation adopts CEFIoT layered architecture to achieve the fault tolerance capabilities and is described in detail in~\cite{Aaltodoc:AsadThesis}. This article presents a high-level overview.
	
\begin{figure}[!htb]
		\centering \includegraphics[scale=0.5]{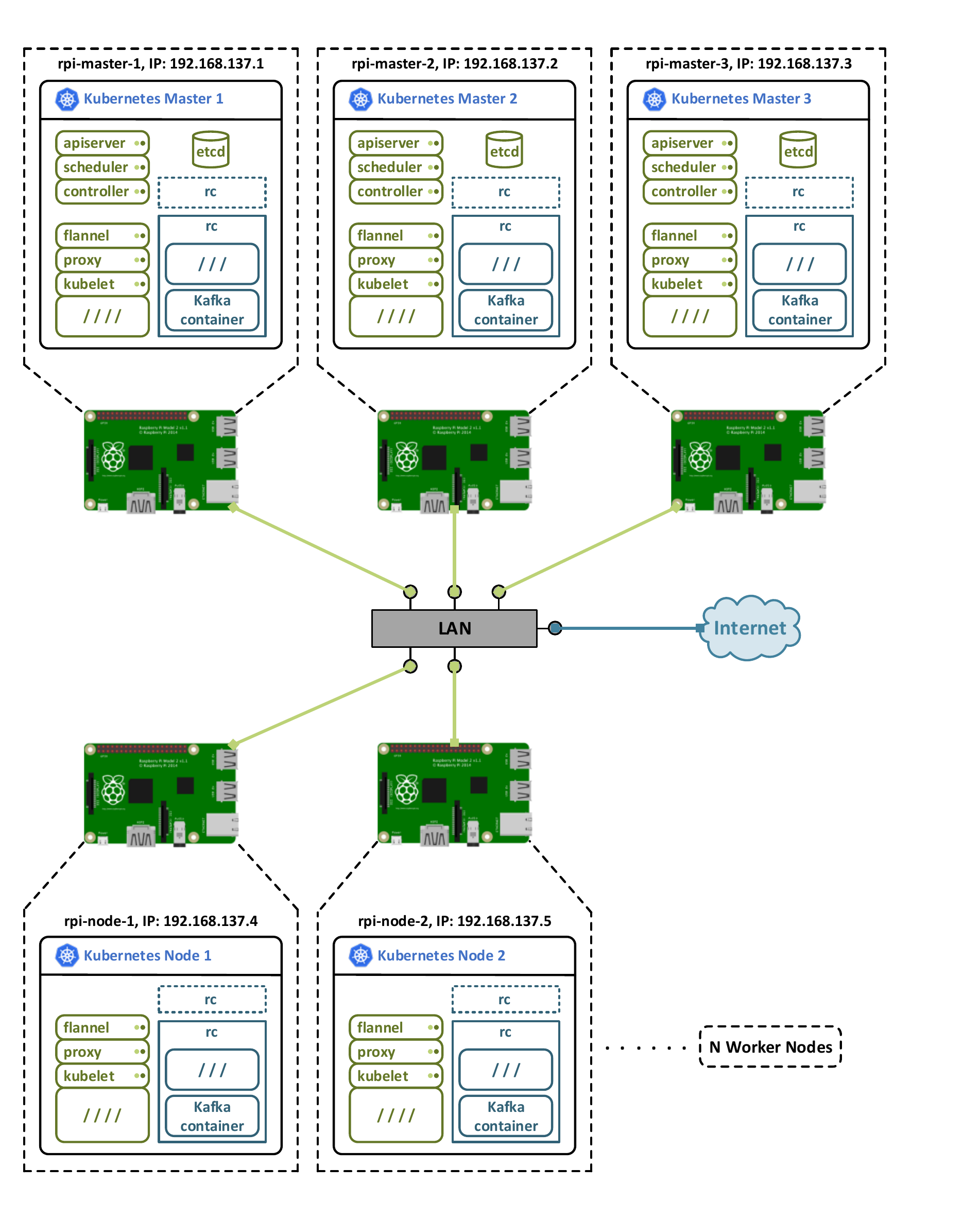}
		\caption{Model of surveillance camera use case, $rc$ is replication controller}
		\label{fig:overview}
	\end{figure}

	Fig.~\ref{fig:overview} shows a running implementation of the selected use case in which five RPi boards are connected to a local area network. Each RPi runs specific processes as Docker containers that communicate through Kafka and are managed by Kubernetes. As can be seen, there are three master nodes (\textit{rpi-master-1, rpi-master-2, rpi-master-3}) and two worker nodes (\textit{rpi-node-1, rpi-node-2}) in our implementation. More master and worker nodes can be added (see Fig.~\ref{fig:overview} with a letter~N) to the system. The purpose of configuring three masters is to achieve single node fault tolerance, thus if one master fails, another node will start operating as a new master node.
	
	\subsection{Application Isolation layer: Configure separate containers}
	
	In Fig. \ref{fig:overview}, the first layer is logically observed in the form of various processes that are encapsulated inside Docker containers, where green boxes relate to Kubernetes cluster configuration and the processes in blue correspond to a Kafka cluster. The fault-tolerant configuration database \textit{etcd} is configured on master nodes as a cluster of three \textit{etcd} servers, which ensure the availability of Kubernetes cluster management in case of a node failure. This service notifies other components when events such as configuration data changes or node failures happen. \textit{Flannel} provides container-to-container communication by assigning unique IP addresses. \textit{Kubelet} starts three processes named apiserver, controller manager, and scheduler that are used by the master nodes to perform container management. Last but not least, the \textit{Proxy} service forwards requests to the correct container in addition to working with \textit{etcd} to provide master node fail-over.
	
	\subsection{Data Transport layer: Kafka cluster framework}
	
	This layer is modeled in terms of data pipeline that provides data transport through Kafka pub/sub cluster. Before transmission, the data (motion-detected images) are collected using the external camera sensors that are attached to each RPi node. Once the data have been gathered and processed, they are sent to Kafka topics using separate containers which are created by utilizing the replication controller (rc) feature of Kubernetes. Inside every container, a volume directory is mounted to the local directory of RPi where images are stored. Kafka producer (container) takes an image from the directory, performs processing, and sends it to the local topic. Once the image has been sent, it is received by the Kafka consumer which then sends the image to the remote cloud topic.
	
	\subsection{Multi-cluster Management layer: Kubernetes orchestration}
	
	In Fig.~\ref{fig:overview}, the third layer is logically mapped to the management and monitoring of containers in this surveillance camera system. In this layer, the data computation is placed physically on multiple nodes. When the data are in the process of transportation, they are stored locally. These data are then consumed inside a separate container which runs on any of the five system nodes. That container is automatically started using the \textit{rc} feature of Kubernetes. If the scenarios of network connectivity outage or the malicious harming of the edge node materialize where the container was running, Kubernetes resumes the consumer process on another node, keeping the system available and data processing pipeline active. 
	
	The architectural design of the surveillance camera system is analyzed in terms of various fault tolerance scenarios (for details, see \cite{Aaltodoc:AsadThesis}). It is observed that the system is able to tolerate a two-node failure in the cluster of five nodes. In case of a smaller cluster having three nodes, it is still able to tolerate one single node failure. Consider an example scenario in which there is a camera sensor which streams and processes live video feed with motion detected in it from one of the system nodes. During transmission, a node is physically damaged by some criminal activity which stops data stream to be processed on that particular node. However, the data that have already been replicated in the cluster and stored simultaneously to the cloud back-end will remain and be retrieved from other system nodes. Therefore, the CEFIoT architecture is able to address the fault tolerance requirements of our application.
	
	\section{Conclusion and Future Work}\label{sec:conclusion}
	In this paper, a new fault-tolerant architecture CEFIoT has been proposed for IoT applications. It addresses the following problems: (\rmnum{1})~the processing stages in IoT applications need to have separate implementations for both the edge and the cloud, (\rmnum{2})~the placement of computation is inflexible with separate software stacks, as the optimal deployment decisions need to be made at runtime, and (\rmnum{3})~unified fault tolerance is essential in case of intermittent long-distance network connectivity problems, malicious harming of edge devices, or harsh environments. The architecture is based on state-of-the-art cloud technologies including Docker, Kubernetes, and the Apache Kafka framework and deploying them also for edge computing. The architecture has four features: (\rmnum{1}) it overcomes the physical node failure and long-distance network problems by providing replication-based local fault tolerance on edge devices, (\rmnum{2}) it performs processing at the edge in order to minimize physical distance delay and to consume a smaller amount of network bandwidth, (\rmnum{3}) it allows computation to be moved between the edge and the cloud without any source code modifications, and (\rmnum{4}) it provides a common software stack for computing, portability, and management ease. 
	
	The CEFIoT architecture consists of three layers: (\rmnum{1})~Application Isolation, (\rmnum{2})~Data Transport, and (\rmnum{3})~Multi-cluster Management layer. Based on this layered design, the fault tolerance capabilities have been demonstrated by implementing a surveillance camera use case as an example of a generic IoT application. The case study implementation involves a clustered system of five RPi nodes, which operates by collecting images (with motion detected in them) from external camera sensors attached to each RPi node, replicating them across the local edge cluster, and sending them to the cloud back-end for further processing and storage. It is observed that the selected use case satisfies the key design features of CEFIoT architecture through functional evaluation \cite{Aaltodoc:AsadThesis}.   
	
	In the future, we will extend the current work to: (\rmnum{1})~Management and monitoring system for automatic software updates across IoT architectures (similar to the security and deployment techniques of resin.io platform) and (\rmnum{2})~Integration with the federated cluster management of Kubernetes in order to have single management system for a large number of both edge and cloud clusters. 
	
	\bibliographystyle{IEEEtran}
	\bibliography{references-new}
	
\end{document}